# Atomistic Modelling of High-Entropy Layered Anodes and Their Electrolyte Interface

Amreen Bano* and Dan T Major

Department of Chemistry and Institute of Nanotechnology and Advanced Materials, Bar-Ilan University, Ramat Gan, Israel-5290002

*Email: banoamreen.7@gmail.com

**Abstract**

Van der Waals (vdW) heterostructures have attracted intense interest worldwide as they offer several routes to design materials with novel features and wide-ranging applications. Unfortunately, at present, vdW heterostructures are restricted to a small number of stackable layers, due to the weak vdW forces holding adjacent layers together. In this work, we report on computational studies of a bulk vdW material consisting of alternating $TiS_2$ and $TiSe_2$ (TSS) vertically arranged layers as a potential candidate for anode applications. We use density functional theory (DFT) calculations and ab-initio molecular dynamics (AIMD) simulations to explore the effect of high entropy on several electrochemically relevant properties of the bulk heterostructure (TSS-HS) by substituting $Mo^{6+}$ and $Al^{3+}$ at the transition metal site ($Ti^{4+}$). We also study the solvation shell formation at the electrode-electrolyte interface (EEI) using AIMD to determine Li-coordination. Based on the properties computed using DFT and AIMD we propose that high entropy TSS-HS (TSS-HE) might possess improved electrochemical performance over standard TSS-HS. Factors that could improve the performance of TSS-HE are 1) Less structural deformation, 2) Strong bonding (Metal-Oxygen), 3) Better electron mobility, 4) Wider operational voltage window, and 5) Faster Li-ion diffusion. Our observations suggest that 'high entropy' can be an effective strategy to design new anode materials for improving electrochemical performance of Li-ion batteries.

## Introduction

Owing to their high energy density, low cost, large storage capacity and cycling stability, lithium-ion batteries (LIBs) have gained tremendous attention as electronic energy storage devices.[1-4] Since the discovery of graphene, with its high electroconductivity and specific surface area, graphene-based anode materials have been studied extensively.[5] However, poor mechanical stability and low specific capacity[6] are some limiting factors which are difficult to overcome, prompting researchers to look for improved or alternative anode materials. These strategies include doping,[7, 8] strain,[9, 10] defects,[11, 12] or heterostructures with different layered materials.[5, 13] In this regard, heterostructures are an attractive alternative that have already shown great promise. The distinct layers of heterostructures are held together by interlayer van der Waals (vdW) interactions. These heterostructures not only exhibit new basic physics,[14, 15] but have also shown considerable performance in applications such as field-effect transistors[13, 16, 17] and photodetectors.[18-20] Additionally, there is a plethora of literature focusing on the virtues of graphene-like heterostructures used as electrode materials.[21-23] Some promising post-graphene candidates are transition metal dichalcogenides (TMDs),[24-27] phosphorene,[28] and silicene.[29] The layered TMD materials not only have properties similar to graphene, but they also have many mechanical, optical, chemical, thermal, and electrical capabilities that are comparable to or superior to graphene.[27, 30, 31] The weak interlayer dispersion interactions in TMDs or their heterostructures facilitate intercalation of foreign atoms (such as Li-ions) which is reported to improve the electron conductivity of the material.[32] Hence, layered TMDs and their heterostructures have previously been proposed as promising anode materials for LIBs.[33, 34]

Commonly, a vdW heterostructure is prepared by mechanically stacking one two-dimensional (2D) layer on top of another. Interestingly, some bulk vdW heterostructures exist in nature, like Frankencite, which is a naturally occurring vdW heterostructure composed of $SnS_2$ and PbS layers stacked alternately.[35] There are few examples of such bulk vdW heterostructures, and synthesizing such heterostructures remains difficult despite the advancements in 2D materials research. Another example of a bulk vdW heterostructure is the 6R-phase of $TaS_2$, which consists of alternating layers of 1H-(superconducting)[36, 37] and 1T-$TaS_2$ (Mott insulator)[38]. Two typical layered TMD materials, $TiS_2$ and $TiSe_2$, have gained traction as potential anode materials.[39-43]

The performance of heterostructures as anode materials has been studied extensively for materials like $MoS_2$/$WS_2$,[44] $WS_2$/$NbSe_2$,[45] and $NiSe_2$/$SnSe_2$.[46] Still, to increase the choice of possible anode materials it is desirable to go beyond these base materials and pursue more complex materials. Considering the difficulty in synthesizing heterostructures it is of value to first consider potential materials in silico. Here we focus on the heterostructures with 'high entropy' elemental configuration. The study of high entropy battery materials commenced with the development of high entropy metal oxides as LIB anodes that showed significant improvement in specific capacity and capacity retention.[47-49] This direction was further expanded to high entropy Li-ion cathodes that showed improved electrochemical performance.[50, 51] We note that the terms 'high entropy',

'compositionally complex' and 'multi-component' are not necessarily interchangeable. Under ideal mixing conditions, where components are randomly distributed, the configurational entropy is expressed as:[52]

$$S_{Config}^{Ideal} = -k_B \sum_{i=1}^{N} x_i \ln x_i \qquad (1)$$

where $x_i$ indicates the mole fraction of component $i$ on the site of mixing. To be considered high entropy, a material must have an $S_{Config}^{Ideal}$ value of $\geq 1.5R$.[53] For instance, the traditional cathode material $LiNi_{0.5}Mn_{0.3}Co_{0.2}O_2$, has an $S_{Config}^{Ideal}$ of $1.03k_B$ per transition metal, and therefore cannot be considered a 'high entropy' material.[54] High-entropy materials can produce a large ensemble of local environments instead of the formation of local homogenous domains with clusters of like atoms which can form in low-entropy materials. Studies have indicated that multi-ensemble local environments might reduce the short-range ordering which is one of the causes of impaired electrochemical performance. Reduced short-range ordering enables 'high entropy' materials to attain higher rates and capacities.[51, 55] In addition, compared to low-entropy materials, high-entropy materials show a greater tolerance for lattice distortions, allowing for significant changes to the energy landscape for ion diffusion. As recently demonstrated in high-entropy oxide-based materials, carefully designed lattice distortions can produce percolating diffusion routes, allowing orders of magnitude increase in ionic conductivities.[56]

In this work, we studied the effect of high entropy on structural, electronic, bond strength, and electrochemical characteristics of a typical bulk heterostructure anode TSS-HS using DFT and AIMD calculations. Additionally, we explored Li-ion diffusion at the anode-electrolyte interface. Our analysis suggests that high entropy TSS-HS anode materials can improve overall electrochemical performance and warrant further experimental study.

**Computational Methods**

Using DFT, we performed high-throughput calculations on pristine and high entropy 1T-$TiS_2$-$TiSe_2$ (TSS) bulk heterostructure (HS) to explore its potential as a layered anode material for Li-ion batteries. To meet the 'high entropy' criteria, i.e., $N \geq 5$, we substitute Mo and Al at the transition metal site (Ti). Several Ti-substitution sites were explored using DFT to find suitable spots for Mo and Al in TSS-HS. Using *equation (1)*, we obtained $S_{Config}^{Ideal}=1.44k_B$ per component for our TSS-HE anode model. The optimized TSS-HS and high-entropy TSS (TSS-HE) are shown in Fig. 1 (a, b). A supercell of 3×3×2 was used to build the TSS-HS and TSS-HE models. All DFT calculations used the Vienna ab initio simulation package (VASP).[57, 58] For all computations, projector augmented wave (PAW) potentials[59] were utilized. The electron-exchange correlation term was treated using the Perdew, Burke, and Ernzerhof (PBE) functionals which is based on the generalized gradient approximation (GGA).[60] We included onsite Coulomb interactions with Hubbard-$U$ terms[61] (PBE+$U$) for strongly correlated elements with $d$-orbitals, such as Ti and Mo.

The effective *U* parameters for Ti and Mo used were 4.0 eV[62] and 5.0 eV[63], respectively. Spin-polarized calculations were carried out with a 5×5×2 *k-mesh*. Grimme's D3 dispersion correction [64] was used to treat van der Waals interactions. For the analysis of delithiated states of TSS-HS and TSS-HE, Monte Carlo simulations were performed to identify the preferred Li-ion sites at various Li-concentrations (that is, for lithiated to fully delithiated states: 100.0, 87.5, 75.0, 67.5, 50.0, 32.5, 25.0, 12.5, 0.0). Using the Ising model, we generated 20,000 initial guess ground state magnetic configurations for the TSS-HS and TSS-HE systems for the DFT calculations. The lowest three of the 20,000 potential spin configurations were taken into consideration for DFT calculations. The lowest energy structure obtained using DFT was considered for final calculations and analysis. We found that TSS-HS prefers a ferromagnetic state, while TSS-HE prefers an antiferromagnetic state as the ground state magnetic ordering. For the plane-wave basis set, we employed a 520 eV kinetic energy cut-off. In geometry optimizations, the force per atom convergence criterion was 0.01 eV/Å, whereas the convergence criterion for self-consistent field electronic structure computations was $1\times10^{-5}$ eV.

To determine the strength of metal-anion bonds at various Li-concentrations (metal=Ti, Mo and Al; anion=S and Se), we performed partial crystal orbital Hamiltonian population (pCOHP) analysis using the LOBSTER package.[65-68] Maximum bond distances between Ti-S/Se, Mo-S, and Al-Se were between 1-3 Å for the pCOHP calculations. This analysis provides clues regarding the 'high entropy' effect on structural deformation of the TSS under the effect of Li-extraction.

Furthermore, to obtain the voltage profile of Li-(de)intercalation of TSS-HS and TSS-HE, the following expression was used:[69-71]

$$V = - \frac{[E(Li_{x+dx}TSS) - E(Li_xTSS)]}{dx} + E(Li_{bcc}) \qquad (2)$$

Here, $E(Li_{x+dx}TSS)$ and $E(Li_xTSS)$ are the energies per formula unit of TSS-HS (or TSS-HE) at $x + dx$ and $x$ Li-concentrations. $E(Li_{bcc})$ is the energy per formula unit of bulk Li metal.

We also explored the relative performance of TSS-HS and TSS-HE anode materials forming an interface with a basic, common electrolyte composed of EC:PC (1:1) organic solvent molecules with 1M $LiPF_6^-$ salt.[72, 73] To analyze the formation of a solid-electrolyte interface (SEI) between the anode (TSS-HS/TSS-HE) and the electrolyte we used ab-initio molecular dynamics simulations (AIMD) as implemented in VASP [57, 58] at a temperature of 300K. These simulations demand significant computational resources, and therefore we used a reduced 3×3×2 k-mesh.[74] The canonical ensemble (NVT) with the Nose-Hoover thermostat as a heat bath was employed. A time step of 1.0 *fs* and a total of 4 *ps* simulation time was considered. The diffusion coefficient of Li-ions was determined using mean square displacement (MSD) as follows: [75, 76]

$$MSD\ (t) = \langle [r(t) - r(0)]^2 \rangle \qquad (3)$$

Employing the Einstein formula, we obtain the diffusion coefficient as follows:

$$D = \frac{1}{6}\frac{d\,MSD(t)}{dt} \qquad (4)$$

We note that these equations are only approximate single particle diffusion equations rather than cooperative multi-ion diffusion equations. We used the Packmol software [77] to create the optimized anode-electrolyte interface. The electrode-electrolyte interface modelled via AIMD simulations is shown in Fig 1(c).

## Results and Discussion

1. Bulk Properties
   *1.1 Structural Stability*
   
   The pristine and high-entropy bulk HS composed of alternating layers of TiS$_2$ and TiSe$_2$ are shown is Fig 1 (a, b). The initial lattice parameters were chosen from the parent layered material TiSe$_2$,[78] which were then allowed to fully relax to obtain the structural parameters of the complete HS. The lattice parameters obtained for the TSS-HS and TSS-HE systems are listed Table 1.

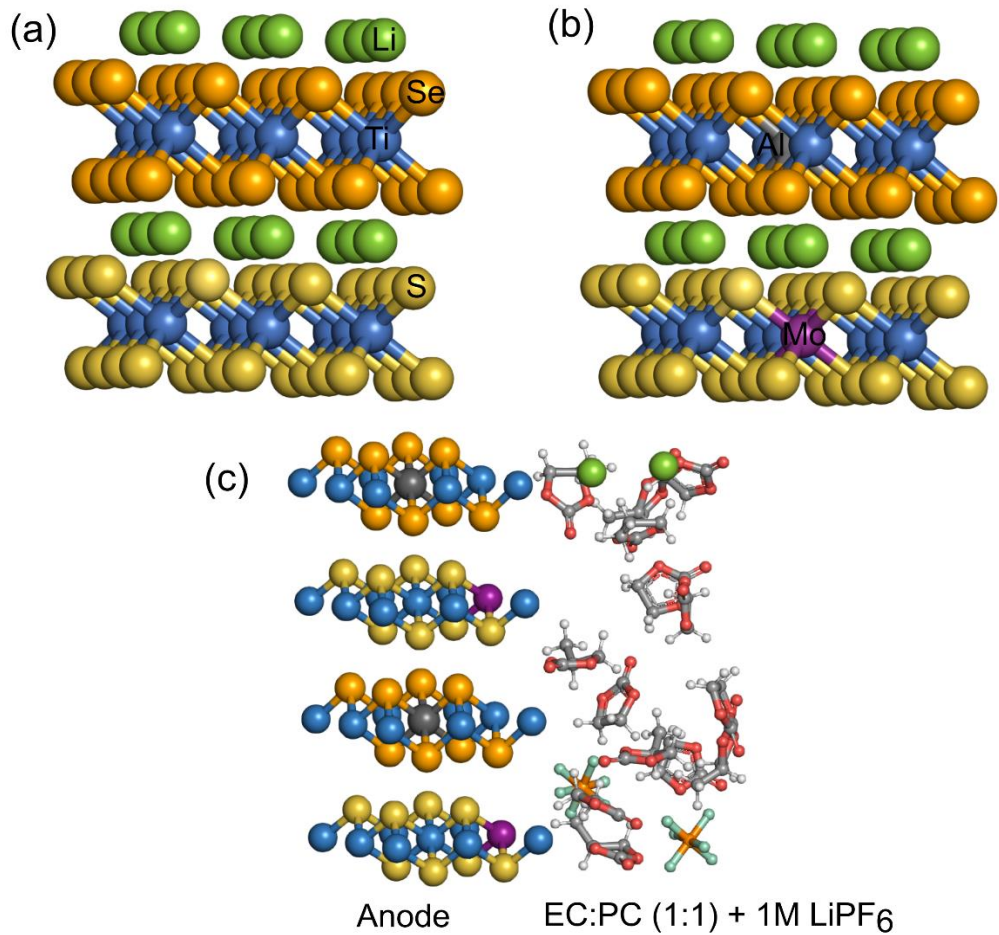

Figure 1: Optimized crystal structures of (a) TSS-HS and (b) TSS-HE. (c) Electrode-electrolyte interface obtained using Packmol.[77]

These parameters are obtained without intercalating any Li-ion within the layers of the HS. The overall lattice mismatch obtained for TSS-HS and TSS-HE *w.r.t* the parent material TiSe$_2$ lattice parameter *'a'* is 0.28% and 2.52%, respectively, which is quite low and considered within the limits of minor mismatch.[79]

Table 1: DFT calculated lattice parameters using PBE-U-D3 for TSS-HS, TSS-HE, and the reference heterostructure material TiSe$_2$.

| Structure | *'a'* (Å) | *'c'* (Å) | *volume* (Å$^3$) |
|---|---|---|---|
| TSS-HS | 3.58 | 5.84 | 53.68 |
| TSS-HE | 3.66 | 5.87 | 56.47 |
| TiSe$_2$ | 3.57 [78], 3.533 [80] | 5.995 [80] | 69.54 |

To identify low energy sites for Li-ions we calculated the ground state energy of four different available intercalation sites as shown in Fig 2 (inset). The lowest energy was obtained at site-1 (i.e., atop the Ti-site, $E_1$) and the energies for all other sites are provided in Fig 2 using the energy of site-1 as a reference. All other sites have relative energies, $\Delta E \gtrsim 1.0$ eV. For further calculations site-1 was considered for Li-intercalation, and 100% Li-coverage was used for fully lithiated layered anode materials.

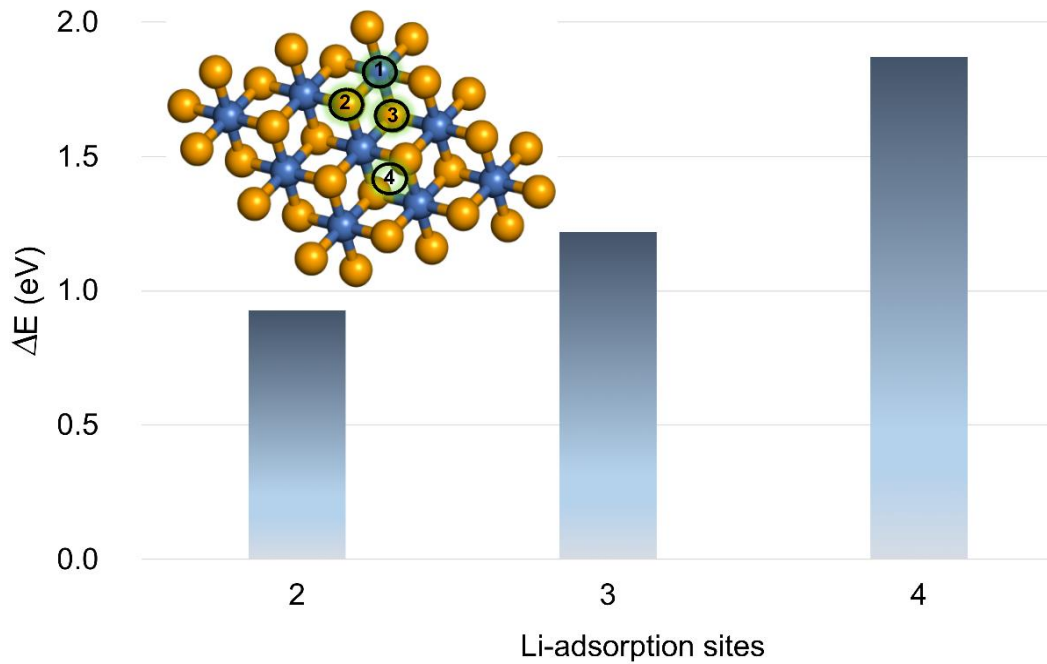

Figure 2: The difference in ground state energies of all considered sites relative to site-1. Li-intercalation sites labeled 1, 2, 3, and 4 are displayed along the 'c'-axis(inset).

Furthermore, in order model the effect of 'high-entropy' on the structural changes during cycling of the layered anode TSS-HS, the relative changes in structural parameters *'a'*, *'c'* and *volume* were calculated (Fig 3). In Fig 3a, *'a'* exhibits a pronounced decreasing trend with lithiation (that is, charging) up to 50% Li-concentration in TSS-HE, while subsequent lithiation (Li>50%) results in a smaller reduction in *'a'*. In TSS-HS, *'a'* shows a continuous, albeit somewhat erratic, shrinkage as a function of Li-concentration, which overall is greater than for the high entropy material. We ascribe this shrinkage to reduced intralayer repulsion between metal ions as they are reduced. The *'c'* parameter changes only slightly as a function of lithiation up to 75%, while there is a drastic reduction upon complete lithiation for both TSS-HS and TSS-HE, which is also mirrored in the volume behavior. We ascribe this shrinkage to reduced interlayer repulsion between Se and S because of Li-ion intercalation between the $TiS_2$ and $TiSe_2$ layers. We further note that the overall change in the *'c'* parameter and volume is less for the high entropy material.

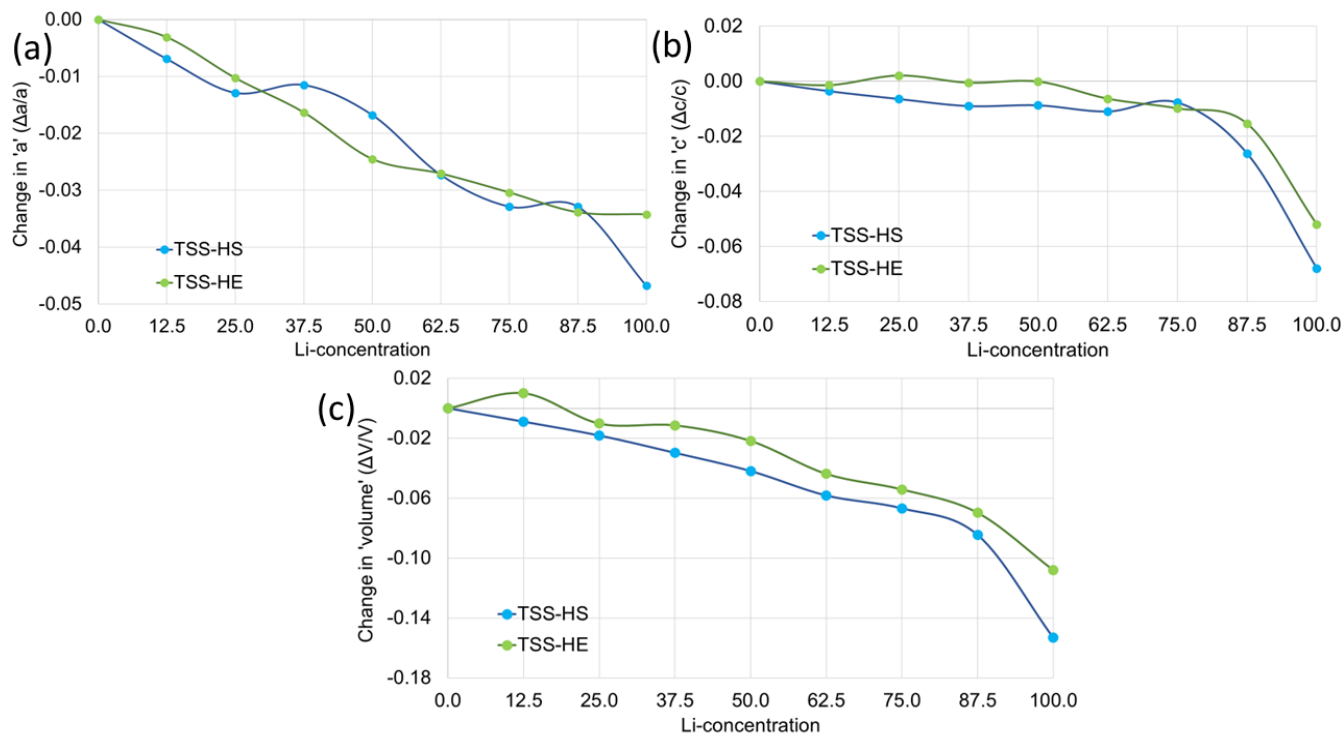

Figure 3: Relative change in (a) *'a'* lattice parameter, (b) *'c'* parameter and (c) *volume* of TSS-HS and TSS-HE as a function of Li-concentration.

Similar trends were observed in *volume* change as well (Fig 3c). Suppressed structural deformation attributed to 'entropy-stabilization' during dis-charging has been reported by several groups.[50, 81] Low lattice deformation may offer enhanced specific capacity as well as prolonged electrochemical cycling.

2. *Electronic Structure*

We performed electronic structure calculations to analyze the effect of 'high-entropy' on the electronic structure and electron-mobility in TSS anode materials. Partial density of states (PDOS) obtained for HS and HE at different Li-concentrations are provided in Fig 4, S1 and Fig 5, S2, respectively. Initially at Li-0%, TSS-HS have $Ti^{4+}$ ions which consists of *zero* unpaired *d*-electrons, thus contributing no magnetic moment. This can also be observed in the Li-0% PDOS, where the spin-up and spin-down channels have identically occupied energy levels. The majority of the $Ti^{4+}$-*d* states are found in the conduction band region and no-energy gap is present. Upon lithiation of TSS-HS, the Ti-ions commence redox-activity, that is $Ti^{4+}$ ions are reduced to $Ti^{3+}$ ions. Ultimately all $Ti^{4+}$-ions are reduced to $Ti^{3+}$ ions at the Li-100% level. We see that the $Ti^{4+}$-*d* states are shifted down in energy upon Li-insertion as these ions are being reduced. The $Ti^{3+}$ ion have an unpaired electron, and hence a *non-zero* magnetic moment is obtained for the $Ti^{3+}$-*d* states, which are located across the energy spectrum. As expected, Se-*p* states are observed near the Fermi level, somewhat higher than the S-*p* states. No significant change in the chalcogen ions electronic states is observed with lithiation in TSS-HS, suggestion no detectable anion redox. At the 100% Li-level, a band gap of ~0.32 eV is observed, which may impede electron mobility in deep-charged states of the anode.

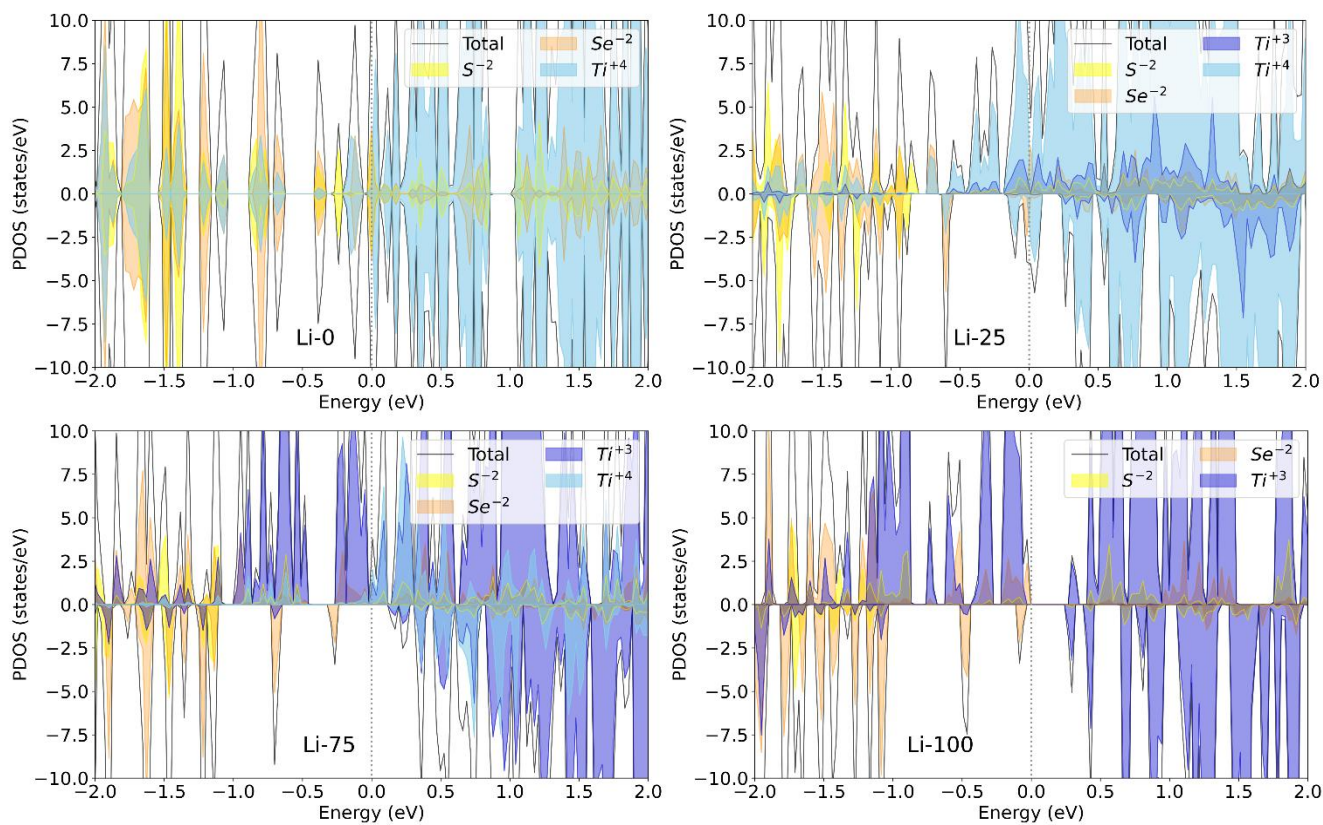

Figure 4: Partial density of states of TSS-HS at different Li-concentrations (Li-0 to 100%). The dotted line indicates that the Fermi level was set at 0.0 eV.

The PDOS of the high-entropy anode TSS-HE (Fig 5, Fig S2) shows similar trends for the $Ti^{4+} \rightarrow Ti^{3+}$ reduction with lithiation. The dopants $Mo^{6+}$ and $Al^{3+}$ were not found to change oxidation states during lithiation, suggesting no direct participation in redox, similar to what has been observed for cathode doping and high entropy materials. [50, 63, 82, 83] However, $Mo^{6+}$ has high-spin states with *non-zero* magnetic moment, whereas the electronic states of $Al^{3+}$ are not spin polarized.

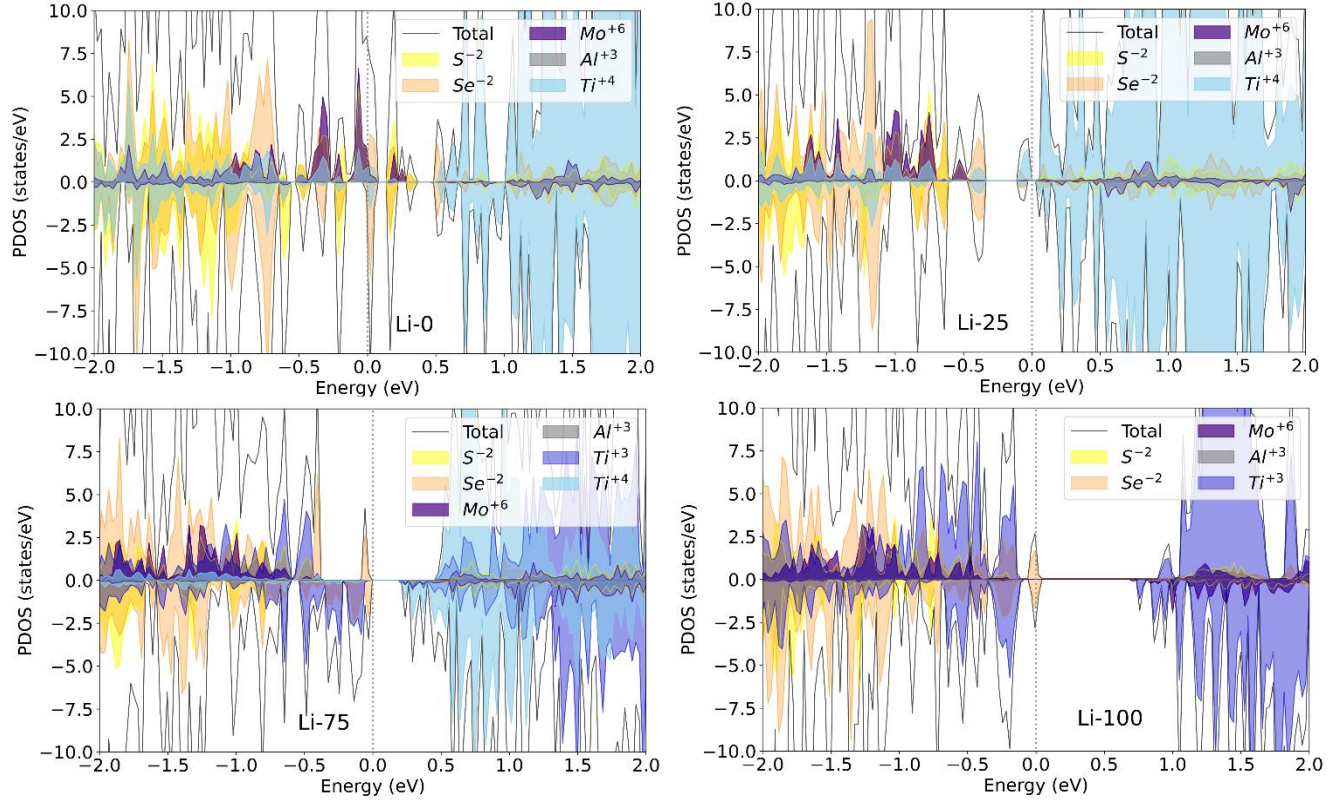

Figure 5: Partial density of states of TSS-HE at different Li-concentrations (Li-0 to 100%). The dotted line indicates that the Fermi level was set at 0.0 eV.

The high-valent dopant (i.e., $Mo^{6+}$) in TSS-HE adds conductive states into the Fermi level region, resulting in a zero-band gap at the 100% Li-level. This may offer improved electron mobility in deep charged states resulting in better conductivity.

*3. Redox Activity and Bond Strength*

In the PDOS analysis, we found that Ti-ions are redox-active, while the chalcogen- and dopant ions do not change their oxidation states during lithiation. For a closer look at the redox activity of the Ti-ions with Li-concentration, we performed redox population analysis for TSS-HS and TSS-HE anode materials. In both TSS-HS and TSS-HE pristine (discharged) states, all Ti-ions are found in a $Ti^{4+}$ oxidation state (Fig 6). In TSS-HS $Ti^{3+}$ begin to appear at Li-25.0%, while in TSS-HE $Ti^{3+}$ emerge at Li-37.5%. In TSS-HS, all $Ti^{4+}$ are consumed by Li-87.5%, while for TSS-HE this only occurs at Li-100.0%. Hence, the onset of Ti-reduction is shifted slightly to greater lithiation levels in the high entropy material. We note that the presence of $Ti^{4+}$ ions is important to maintain a consistent and strong bonding environment with surrounding Se and S ions. Additionally, it has been reported that $Mo^{6+}$ and $Al^{3+}$ ions offer strong bonding with surrounding anions which can lead to improved material stability.[50, 63, 83, 84]

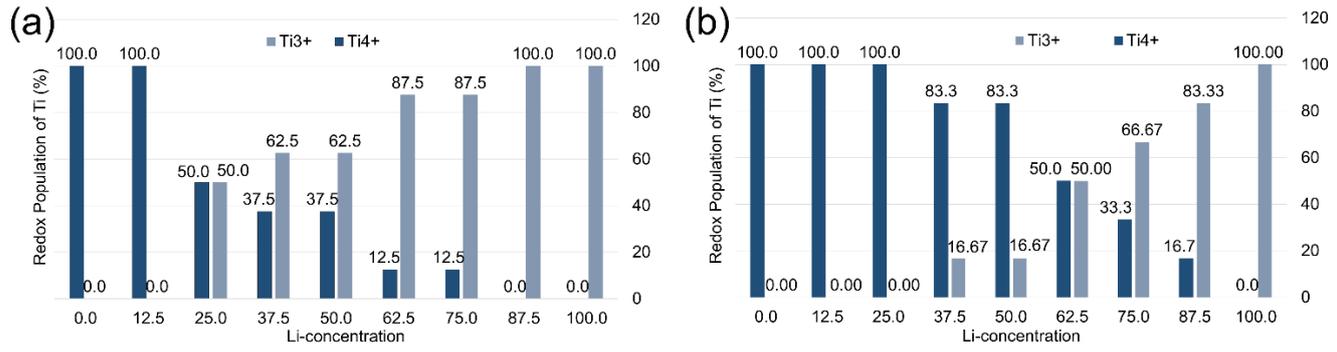

Figure 6: Redox population of Ti-ions as a function of Li-concentration in (a) TSS-HS and (b) TSS-HE.

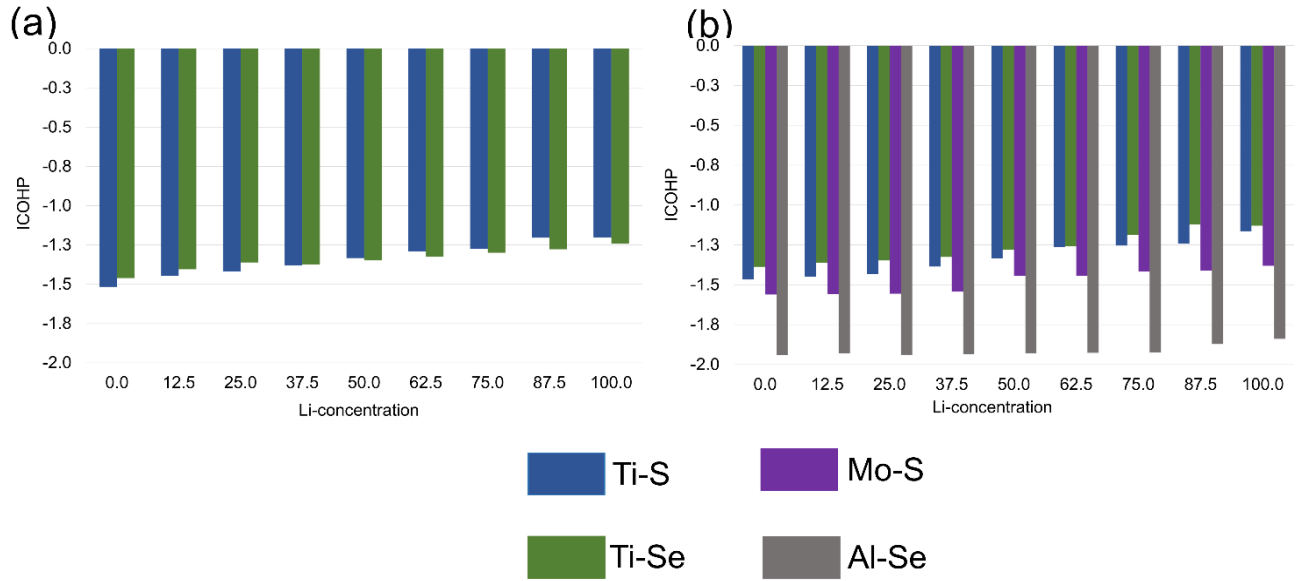

Figure 7: Metal-Chalcogen ion bonding in (a) TSS-HS and (b) TSS-HE, showing the change in bond strength at different Li-concentration levels. Values of ICOHP (y-axis) are obtained as the integrated COHP at the Fermi Level.

To quantify the strength of chalcogen-metal bonds in the TSS materials, we performed integrated COHP (ICOHP) analyses, where more negative ICOHP values indicate stronger bonds. The ICOHP as a function of Li-concentration can be seen in Fig 7 (a) and (b) for TSS-HS and TSS-HE anode materials, respectively. From Fig 7 we observe that Ti-S and Ti-Se bonding is weakened with lithium insertion for both TSS-HS and TSS-HE, which may be attributed to $Ti^{4+} \rightarrow Ti^{3+}$ reduction. However, in Fig 7(b) we see that the dopants $Mo^{6+}$ and $Al^{3+}$ form much stronger bonds with surrounding chalcogen ions. This could be a factor in preventing structural deformations with lithiation in TSS-HE.

## 4. Open Circuit Voltage Profile

In designing high-performance LIBs, open circuit voltage (OCV) is a crucial factor for the development of high energy density batteries. Using equation 2, OCV profiles (Fig 8) were obtained using the ground-state energies of TSS-HE and TSS-HS at each delithiated state.[50]. A wider operational voltage window (~0.47 V) was achieved for TSS-HE compared to TSS-HS (~0.40 V) which may lead to higher capacity as more $Li^+$ ions can be extracted with a larger voltage window. The HE-induced widened

OCV range can be ascribed to improved structural stability accompanied with suppressed volumetric strain in the TSS-HE anode.

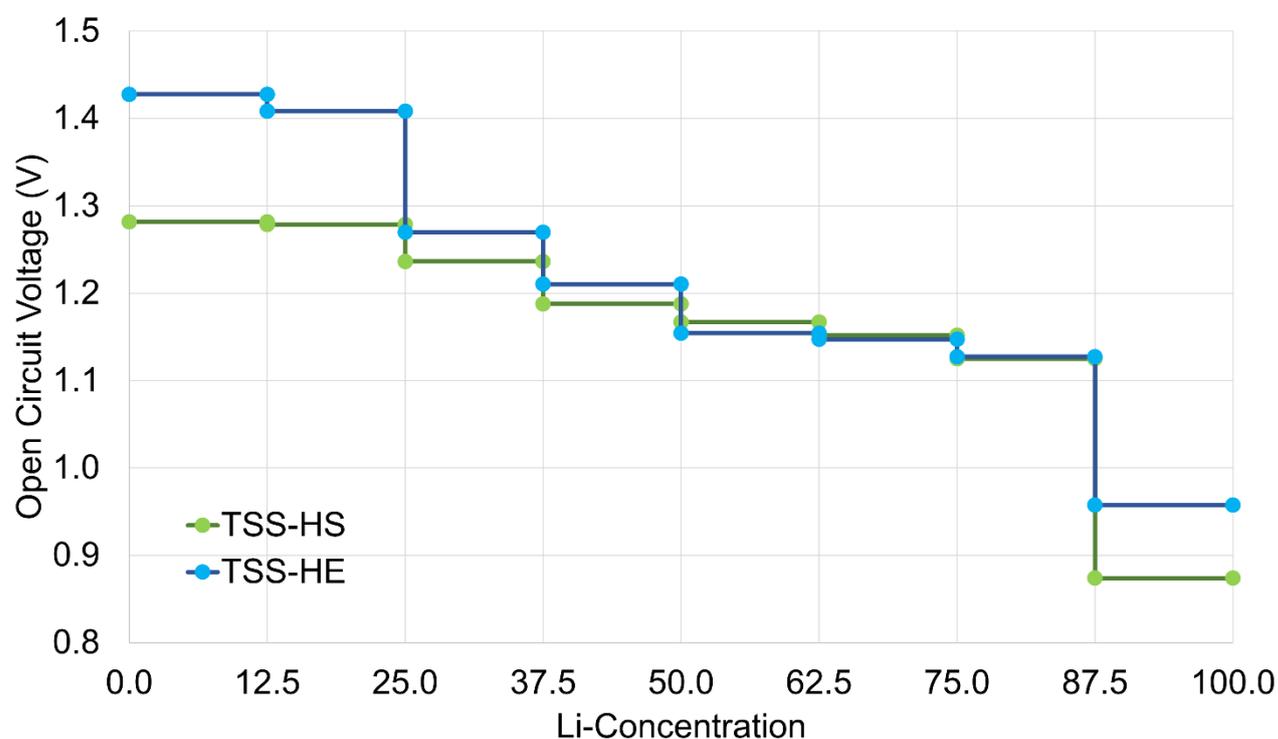

Figure 8: DFT calculated open circuit voltage versus Li-concentration for TSS-HS and TSS-HE.

## 5. Interface properties

Prior to studying the Li diffusion behavior at the EEI, the structural evolution and Li diffusion in a model electrolyte (EC, PC, and $PF_6^-$) was analyzed by computing the integrated radial distribution function (IRDF) (Fig 9(a,b)). We considered relatively short simulations [85-87] to establish basic qualitative trends relating to ion-transport in TSS-HS and TSS-HE at the EEI. Our IRDF analyses show that $Li^+$ ion coordination with surrounding molecules in the electrolyte is slightly reduced in the TSS-HE-electrolyte complex (Fig 10(b)) compared to the TSS-HS-electrolyte (Fig 9(a)), with values of 1.90 to 1.33 (Li-O (EC)) and 1.05 to 0.65 (Li-O (PC)), respectively, within an interaction distance of 3Å. Reduced coordination of Li ions can offer better mobility near the EEI and thus might result in better diffusion. Additionally, we can also predict from the IRDF results that the primary solvation shell at the EEI will mainly be composed of EC and PC molecules.[88]

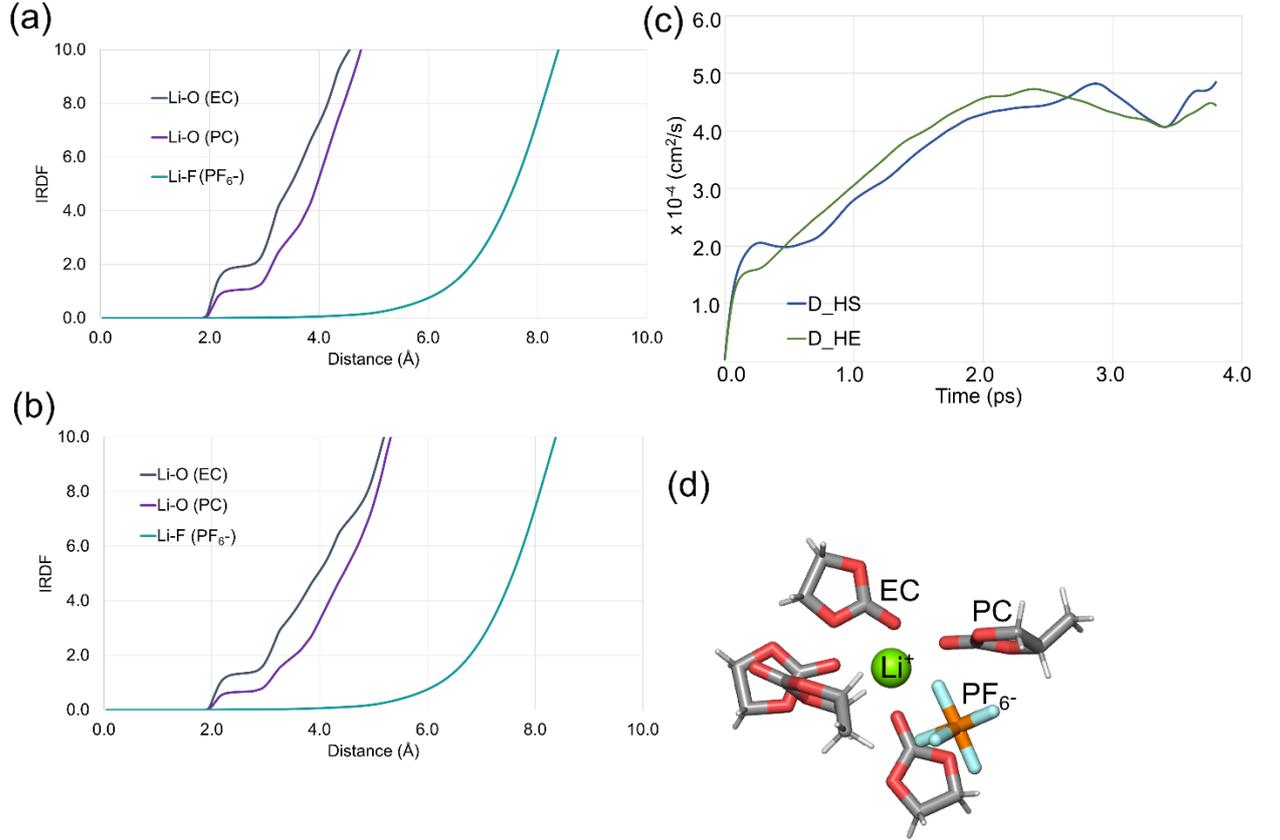

Figure 9: Integrated radial distribution function $n(r)$ obtained with AIMD simulations of Li-ion and its surrounding organic solvents (EC and PC) and salt ($PF_6^-$) for anode materials (a) TSS-HS and (b) TSS-HE. (c) Diffusion coefficients of Li-ions at the EEI interface showing slightly faster diffusion at the TSS-HE:Electrolyte interface (D_HE), compared to the TSS-HS:Electrolyte interface (D_HS), (d) Snapshot of a $Li^+$ ion and its surrounding electrolyte environment within a radius of 4 Å.

The diffusion coefficients of Li ions obtained from a linear fit of the MSD using *equation (4)* are shown in Fig 9(c). We observe that at the EEI with TSS-HE, there is slightly faster Li diffusion compared to EEI with TSS-HS. The average diffusion coefficient and ionic mobility obtained are listed in Table 2.

Table 2: Computed average diffusion coefficient and ionic mobility at 300K.

| EEI | Average Diffusion Coefficient ($\times 10^{-4}$ $cm^2/s$) | Average Ionic Mobility ($\times 10^{-2}$ $cm^2/s/V$) |
|---|---|---|
| TSS-HS: Electrolyte | 4.67±0.067 | 1.81±0.028 |
| TSS-HE: Electrolyte | 5.35±0.095 | 2.10±0.079 |

**Conclusions**

To investigate the effect of "high entropy" on the layered bulk heterostructure anode material TSS, we performed DFT calculations of several properties relevant to electrochemistry. We performed DFT calculations of properties as a function of Li-concentration and propose that TSS-HE anode materials might experience less structural deformations, while possessing better electron mobility, bond strengths, and a wider voltage window than the low-entropy analogue. Specifically, our calculations suggest that the presence of elements like $Mo^{6+}$ and $Al^{3+}$ provide reduced internal strain (less volumetric change), stronger bonds, as well as better electron mobility. We performed AIMD simulations where the IRDF, average diffusion coefficient, and average ionic mobility suggest slightly improved ion-transport for the TSS-HE anode near the EEI. In summary, our work provides atomistic level insights into possible high entropy TSS anode materials for Li-ion batteries and presents ideas for future experimental work.

# Atomistic Modelling of High-Entropy Layered Anodes and Their Electrolyte Interface

Amreen Bano* and Dan T Major

Department of Chemistry and Institute of Nanotechnology and Advanced Materials, Bar-Ilan University, Ramat Gan, Israel-5290002

*Email: banoamreen.7@gmail.com

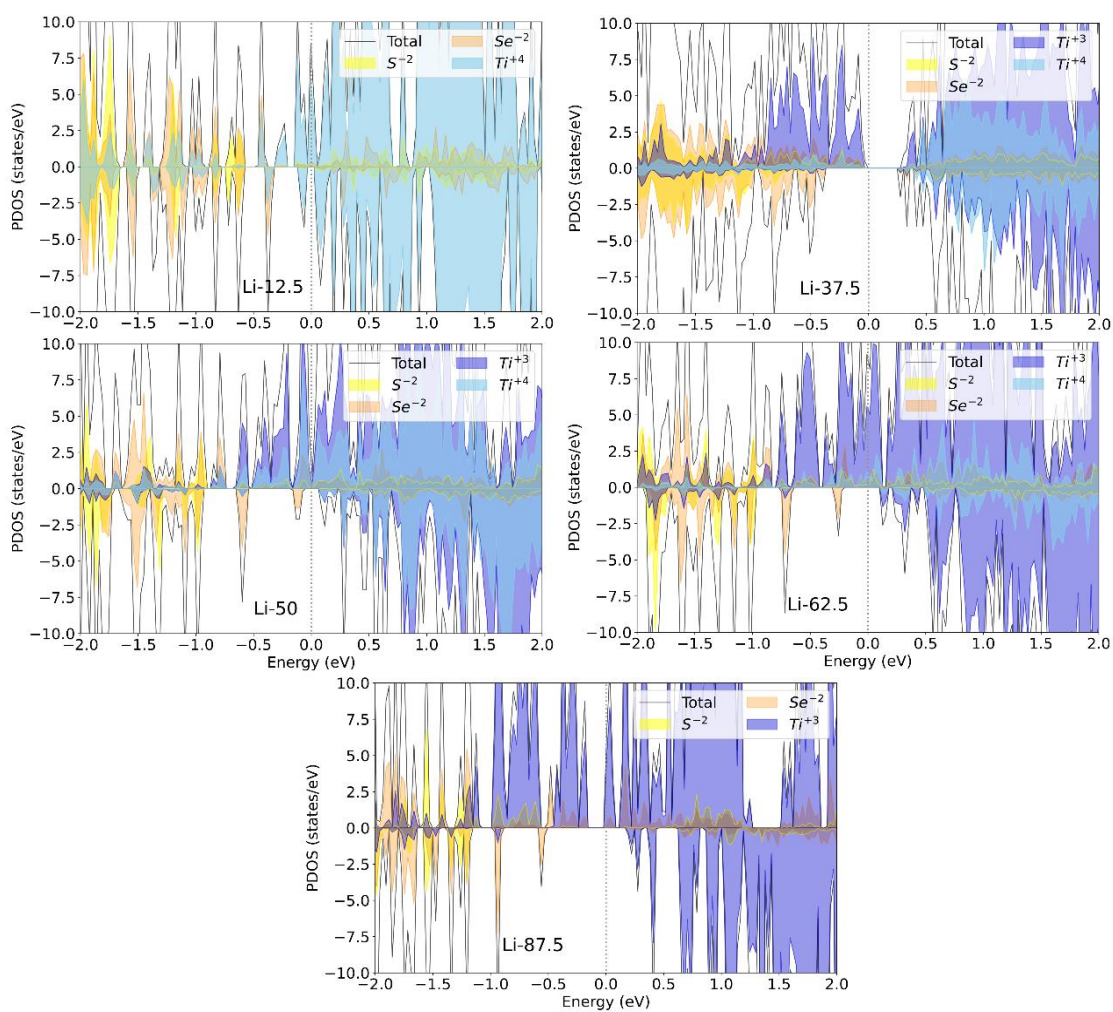

Figure S1: Partial density of states (PDOS) of TSS-HS at different Li-concentration levels showing change in electronic states of constituent elements of the layered anode material. The dotted line indicates that the Fermi level was set at 0.0 eV.

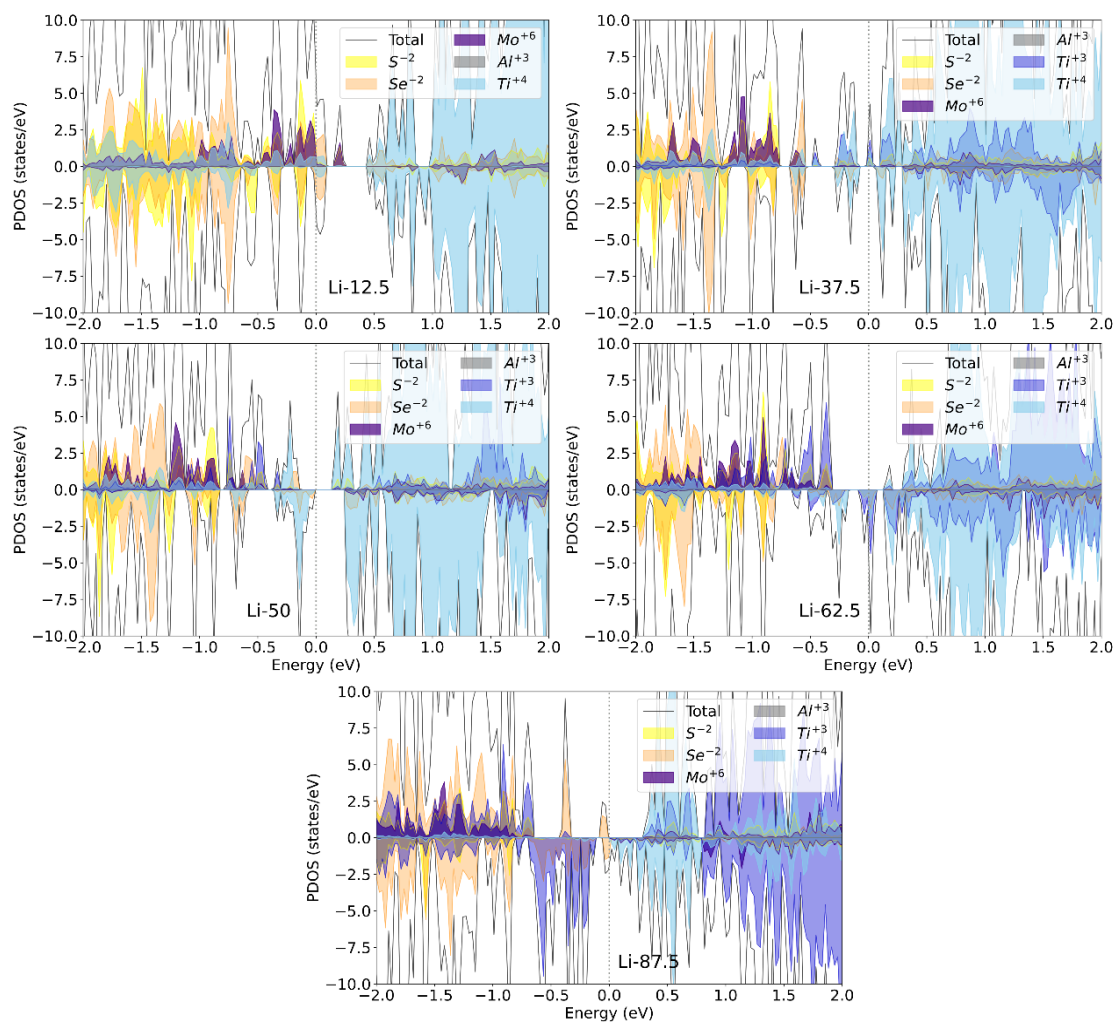

Figure S2: Partial density of states (PDOS) of TSS-HE at different Li-concentration levels showing change in electronic states of constituent elements of the high entropy anode material. The dotted line indicates that the Fermi level was set at 0.0 eV.